\documentstyle[aps,preprint,tighten,epsfig]{revtex}
\def\beq{\begin{equation}}
\def\eeq{\end{equation}}
\def\beqa{\begin{eqnarray}}
\def\eeqa{\end{eqnarray}}
\def\MeV{\nobreak\,\mbox{MeV}}
\def\GeV{\nobreak\,\mbox{GeV}}

\def\pli{p^\prime}
\def\mli{{M^\prime}^2}
\begin{document}
\title{\sc  How Hard Are the Form Factors in Hadronic Vertices with Heavy 
Mesons?}
\author {F.O. Dur\~aes, F.S. Navarra and M. Nielsen \\
\vspace{0.3cm}
{\it Instituto de F\'{\i}sica, Universidade de S\~{a}o Paulo, } \\
{\it C.P. 66318,  05389-970 S\~{a}o Paulo, SP, Brazil}}
\maketitle
\vspace{1cm}

\begin{abstract}
The $ND\Lambda_c$ and  $ND^*\Lambda_c$ form factors are evaluated in a full
QCD sum rule calculation. We study the double Borel sum rule for the three
point function of one meson one nucleon and one $\Lambda_c$ current
up to order six in the operator product expansion. The double Borel transform
is performed with respect to the nucleon and $\Lambda_c$ momenta, and the
form factor is evaluated as a function of the momentum $Q^2$ of the 
heavy meson. These form factors are relevant to evaluate the 
charmonium absorption cross section by hadrons. Our results are compatible
with constant form factors in these vertices.
\\
PACS numbers 14.20.Lq~~12.38.Lg~~14.65.Dw
\\

\end{abstract}

\vspace{1cm}

QCD predicts that at high energy density, hadronic matter will turn into
a plasma of deconfined quarks and gluons, the quark-gluon plasma (QGP), and
there is a great deal of expectation in the comunity that the 
QGP will be observed
in heavy ion collisions at the relativistic heavy ion
collider (RHIC), which just started to operate at the Brookhaven National
Laboratory. As a matter of fact, very recently the NA50 Collaboration reported
\cite{NA50} that the anomalous suppression of $J/\psi$ observed in $Pb+Pb$
collisions at CERN-SPS indicated already the formation of QGP, since these 
experimental data ruled out conventional hadronic models for the $J/\psi$
suppression.

However, there are calculations that reproduce the new NA50 data reasonably
well up to the highest transverse energies \cite{nu01,he02,tho1}, based on
hadronic $J/\psi$ dissociation alone. Therefore, while there are suggestions 
that the anomalous suppression may be due to the formation of the QGP,
other more conventional mechanisms based on $J/\psi$ absorption by comovers
and nucleons have to be still considered. 

The main ingredient in the 
calculations based on hadronic $J/\psi$ dissociation is the magnitude of 
the $J/\psi$ absorption cross section by hadrons, which is not
known experimentally.
In refs.~\cite{nu01,he02,tho1} this cross section was introduced
as a free parameter. Therefore, in order to use the $J/\psi$ suppression
as a signature for the formation of the QGP in heavy ion collisions it is
important to have a better knowledge on the interactions between charmonium 
states and co-moving hadrons.

Various approaches have been used in evaluating the charmonium absorption
cross section by hadrons. Some of them use meson exchange models based
on hadronic effective lagrangians \cite{mamu,ha,ko,tho2,ha2}. The couplings
and form factors needed in these effective lagrangians are not 
phenomenologically known, and sometimes SU(4) relations are used to estimate 
them. However, we don't have any evidence that SU(4) relations should be taken
seriously into account. Also, in general, it is assumed a monopole form factor
at the hadronic vertex \cite{ko,tho2}, with another unknown parameter 
(the cut-off). The results 
obtained for the cross section are very sensitive to the couplings and to the 
form factors \cite{ko,tho2,ha2}. As an example, in ref.~\cite{ko} it was shown
that the $\pi J/\psi\rightarrow DD^*$ cross section may vary by almost one
order of magnitude depending on the value of the  cut-off.
Therefore, it is very important to estimate these 
couplings and form factors with a more theoretical approach. 

In this work we use the QCD sum rule (QCDSR) method \cite{svz} based on
the three-point function to evaluate the 
$ND\Lambda_c$ and $ND^*\Lambda_c$  form factors and coupling constants. In a 
previous work
\cite{nn} we have already evaluated the $ND\Lambda_c$ coupling using
the QCDSR. We have obtained $g_{ND\Lambda_c}=6.7\pm2.1$ while SU(4)
relations used in ref.~\cite{tho2} give $g_{ND\Lambda_c}=14.8$. Of course
the relevance of this difference can not be underestimated since the
cross section is proportional to the square of the coupling constant.
In this work we extend the calculation done before \cite{nn}
and study the $Q^2$ dependence of the form factor for an off-shell
heavy meson, which is the situation relevant for the calculations performed
in refs.~\cite{ha,tho2}.

The three-point function associated with a $ND(D^*)\Lambda_c$ vertex
is constructed with the two baryon currents, $\eta _{\Lambda_c}$ and $\eta _N$,
for $\Lambda_c$ and the nucleon respectively, and the  meson
$D(D^*)$ current, $j_5(j_\mu)$ which we generically call $j_M$: 
\begin{equation}
A_M(p,p^\prime,q)=\int d^4xd^4y \langle 0|T\{\eta _{\Lambda_c}(x)
j_M(y)\overline{\eta }_N(0)\}|0\rangle 
e^{ip^\prime x}e^{-iqy}\; , 
\label{cor}
\end{equation}
with the currents given by \cite{ioffe,dosch}
\beq
{\eta }_{\Lambda _c}= \varepsilon _{abc}({u}_a^TC\gamma _5{d}_b)Q_c \; ,
\eeq
\beq
{\eta }_N= \varepsilon _{abc}({u}_a^TC\gamma^\mu{u}_b)\gamma_5\gamma_\mu d_c 
\; ,
\eeq
\beq
j_5=\overline{Q} i\gamma_5u \; ,
\eeq
\beq
j_\mu=\overline{Q} \gamma_\mu u \; ,
\eeq
where $Q$, $u$ and $d$  are the charm, up and down quark fields
respectively, $C$ is the charge conjugation matrix and $q=p^\prime-p$.

The general expression for the vertex function in Eq.(\ref{cor}) has four
independent structures in the case of $A_5$ \cite{nn} and twelve in
the case of $A_\mu$ \cite{ra}. In principle any of the 
four (twelve) invariant structures appearing in the $A_M$ expression can 
be used to calculate the form factor and the sum rules should yield the 
same result. However, each sum rule could have uncertainties due to the 
truncation in the OPE side and different contributions from the continuum.
Therefore, depending on the Dirac structure we can obtain different results 
due to the uncertainties mentioned above. The tradicional way 
to control these uncertainties, and therefore to check the reliability of 
the sum rule, is to evaluate the stability of the result as a function of the 
Borel mass.

Recently, in ref.\cite{klo} it was pointed out that a better 
determination of  $g_{\pi N N}$ can be done with the 
help of the $\gamma_5\sigma_{\mu\nu}$ structure, since  
this structure is independent of the effective models employed in the 
phenomenological side and it gets a smaller contribution from the
single pole term coming from $N\rightarrow N^*$ transition. This was
confirmed also in the case of the $g_{NKY}$ coupling constant \cite{bnn}. 
Therefore, in this work we will also employ the  $\gamma_5\sigma_{\mu\nu}$ 
structure to evaluate the $ND\Lambda_c$ form factor and coupling constant,
and compare our results with the previous evaluation \cite{nn} carried out 
in the $i\rlap{/}{q}\gamma_5$ structure. In the case of
$ND^*\Lambda_c$ vertex, we will study the sum rule based on the 
$\rlap{/}{p^\prime}\gamma_\mu\rlap{/}{p}$ structure since, in general, 
the structures with a large number of $\gamma$ matrices are more likely to
be stable.

The phenomenological side of the vertex function is obtained
by the consideration of the $\Lambda_c$ and $N$ intermediate states 
contribution to the  matrix element in Eq.(\ref{cor}):
\begin{eqnarray}
A_M^{(phen)}(p,p^\prime,q) &=&\lambda_{\Lambda_c}\lambda_N 
{(\rlap{/}{p^\prime}+M_{\Lambda_c})\over p'^2-
M_{\Lambda_c}^2} g_M(q^2){(\rlap{/}{p}+M_N)\over p^2-M_N^2} + 
\mbox{higher resonances}\; ,
\label{aphen}
\end{eqnarray}
where $\lambda_{\Lambda_c}$ and $\lambda_N$ are
the couplings of the currents with the respective hadronic states and
the meson couplings are given by
\beq
g_5(q^2)=i\gamma_5 {m_D^2 f_D\over m_c}{g_{ND\Lambda_c}(q^2)
\over q^2-m_D^2} \; ,
\label{g5}
\eeq
\beq
g_\mu(q^2)= m_{D^*} f_{D^*}{g_{ND^*\Lambda_c}(q^2)\over q^2-m_{D^*}^2}\left(
-\gamma_\mu + {\rlap{/}{q}q_\mu\over m_{D^*}^2}\right)\; ,
\label{gmu}
\eeq
where $m_D$,  $m_D^*$, $f_D$ and $f_{D^*}$ are the mass and decay 
constant of the mesons $D$ and $D^*$ respectively, and
$m_{c}$ is the $c$ quark mass. 

We will write a sum rule in the structure
$\sigma^{\mu\nu}\gamma_5 p^\prime_\mu p_\nu$ for $g_{ND\Lambda_c}(q^2)$ and
one in the structure $\rlap{/}{p^\prime}\gamma_\mu\rlap{/}{p}$ 
for $g_{ND^*\Lambda_c}(q^2)$ and we call  $F$ the invariant amplitude
associated with these structures. The contribution of higher
resonances and continuum in Eq.~(\ref{aphen})
will be taken into account as usual in the standard form of 
ref.~\cite{io2}.

In the OPE side only even dimension operators contribute to the
chosen structures, since the dimension 
of Eq.(\ref{cor})
is four and $p^\prime p$ takes away two dimensions. The diagrams
that contribute, after a double Borel transformation, up to dimension six 
are shown in Fig.~1. The gluon condensate also contributes, but it always
appears with a large suppression factor which arises from the
two-loop internal momentum integration. Therefore, its contribution
is of little influence and will be neglected. To evaluate 
the perturbative contribution (Fig.~1a) we write a double dispersion relation
to the invariant amplitude, $F$,  over the virtualities $p^2$ and 
${\pli}^2$ holding $Q^2=-q^2$ fixed, and use the Cutkosky's rules
\cite{cut} to evaluate the double discontinuity (see ref.\cite{io2}). After 
doing a double Borel transformation \cite{io2} in both variables
$P^2=-p^2\rightarrow M^2$ and ${P^\prime}^2=-{\pli}^2\rightarrow \mli$, and
subtracting the continuum contribution, we get
\beq
\left[\tilde{F}(M^2,\mli,Q^2)\right]_a=-{1\over4\pi^2}\int_{m_c^2}^{u_0}du
\int_0^{s_0}ds\,\rho(u,s,Q^2)\,e^{-u/\mli}e^{-s/M^2}\; ,
\label{d1a}
\eeq
with
\beqa
\rho(u,s,Q^2)&=&\pm{3\over8\pi^2}{1\over\sqrt{\lambda(s,u,Q^2)}}
\int_0^{s}
dm^2\;\left\{ m^2\left(-1+{m_c^2(s-u-Q^2)+(s-u)^2+Q^2(s+u)\over
\lambda(s,u,Q^2)}\right)\right.
\nonumber \\*[7.2pt]
&+&\left.{2m^4Q^2\over\lambda(s,u,Q^2)}\right\}\Theta(1-(\overline
{\cos\theta_K})^2)\Theta\left(u-Q^2-m_c^2+{Q^2u\over m_c^2}-s\right)\; ,
\label{rho}
\eeqa
where 
\beq
\overline{\cos\theta_K}=2s{u+m^2-m_c^2-p_0^\prime(s+m^2)/\sqrt{s}
\over(s-m^2)\sqrt{\lambda(s,u,Q^2)}}\; ,
\eeq
with $p_0^\prime=(s+u+Q^2)/(2\sqrt{s})$ and $\lambda(s,u,Q^2)=s^2+u^2+Q^4-
2su+2Q^2s+2Q^2u$.

In Eq.(\ref{d1a}) $\tilde{F}$ stands for the double Borel transformation
of the amplitude $F$, and the subscript a refers to the diagram in Fig.~1a.
$u_0$ and $s_0$ give the continuum thresholds for the baryons $\Lambda_c$ and 
nucleon respectively, which are, in general, taken from the mass sum rules.

The next
lowest dimension operator is the quark $c$ mass times the quark condensate 
with dimension four (Fig. 1b). Since we are neglecting the 
light quark masses, only terms proportional to $m_c \langle\overline{q}q
\rangle $ will appear. These terms give, after the double Borel transformation
\beq
\left[\tilde{F}(M^2,\mli,Q^2)\right]_{b}=-{m_c\langle\overline{q}q\rangle
\over4\pi^2}\int_{m_c^2}^{u_0}du\int_0^{s_0}ds\,\alpha(s,u,Q^2)
e^{-u/\mli}e^{-s/M^2}\; ,
\label{d1b}
\eeq
where
\beq
\alpha(s,u,Q^2)=\pm{s(2m_c^2+s-u+Q^2)\over\left(\lambda(s,u,Q^2)\right)^{3/2}}
\Theta\left(u-Q^2-m_c^2+{Q^2u\over m_c^2}-s\right)\; .
\label{alp}
\eeq

The next contribution comes from the diagram involving dimension 6 operator
of the type  $\langle\overline{q}q\overline{q}q\rangle$ ($\simeq \langle
\overline{q}q\rangle ^2$) shown in Fig.~1c:  
\beq
\left[\tilde{F}(M^2,\mli,Q^2)\right]_{c} =\pm  
\frac{\langle\overline{q}q\rangle^2}{3}e^{-m_c^2/\mli} \, .
\label{d1c}
\eeq
In Eqs.~(\ref{rho}) (\ref{alp}) and (\ref{d1c}) the $+$ and $-$  sign refers 
to the $ND\Lambda_c$ and $ND^*\Lambda_c$ vertices repectively.

The Borel transformation of the phenomenological side gives
\beq
\left[\tilde{F}(M^2,\mli,Q^2)\right]_{phen}=\lambda_{\Lambda_c}\lambda_N 
G_{NM\Lambda_c}(Q^2)e^{-M_N^2/M^2}e^{-M_{\Lambda_c}^2/\mli}\;,
\label{fen}
\eeq
where the continuum contribution has 
already been incorporated in the OPE side, through the  continuum thresholds
$s_0$ nd $u_0$. In Eq.~(\ref{fen}) we have defined
\beqa
G_{NM\Lambda_c}(Q^2)&=& { m_D^2 f_D\over m_c}{g_{ND\Lambda_c}(Q^2)\over 
Q^2+m_D^2}\;\;\;{\mbox {for}}\;M=D
\nonumber\\
&=&m_{D^*} f_{D^*}{g_{ND^*\Lambda_c}(Q^2)\over Q^2+m_{D^*}^2}\;\;\;
{\mbox {for}}\;M=D^*\;.
\eeqa

In order to obtain $g_{ND(D^*)\Lambda_c}(Q^2)$ we identify Eq.~(\ref{fen}) 
with the 
sum of Eqs.~(\ref{d1a}), (\ref{d1b}) and (\ref{d1c}). We obtain:
\beqa
g_{ND\Lambda_c}(Q^2)&=&
{e^{M_N^2/M^2}e^{M_{\Lambda_c}^2/\mli}\over\lambda_{\Lambda_c}\lambda_N }{m_c(
Q^2+m_D^2)\over m_D^2 f_D}\left[-{1\over4\pi^2}\int_{m_c^2}^{u_0}du
\int_0^{s_0}ds\,e^{-u/\mli}e^{-s/M^2}\left(\rho(u,s,Q^2)\right.\right. 
\nonumber \\*[7.2pt]
&+&\left.\left.m_c\langle\overline{q}q\rangle\,\alpha(s,u,Q^2)\right)+
\frac{\langle\overline{q}q\rangle^2}{3}e^{-m_c^2/\mli} \right]
\; ,\label{sr}
\eeqa
and
\beqa
g_{ND^*\Lambda_c}(Q^2)&=&
-{e^{M_N^2/M^2}e^{M_{\Lambda_c}^2/\mli}\over\lambda_{\Lambda_c}\lambda_N }{
Q^2+m_{D^*}^2\over m_{D^*} f_{D^*}}\left[-{1\over4\pi^2}\int_{m_c^2}^{u_0}du
\int_0^{s_0}ds\,e^{-u/\mli}e^{-s/M^2}\left(\rho(u,s,Q^2)\right.\right. 
\nonumber \\*[7.2pt]
&+&\left.\left.m_c\langle\overline{q}q\rangle\,\alpha(s,u,Q^2)\right)+
\frac{\langle\overline{q}q\rangle^2}{3}e^{-m_c^2/\mli} \right]
\; ,\label{srs}
\eeqa

For $\lambda_{\Lambda_c}$ and $\lambda_N$ we use the expressions obtained from
the respective mass sum rules for $\Lambda_c$ \cite{dosch} and for the 
nucleon \cite{ioffe,rry}:

\beqa
|\lambda_{\Lambda_c}|^2&=&e^{M_{\Lambda_c}^2/{M'_M}^2}\left\{{m_c^4\over512
\pi^4}
\int_{m_c^2}^{u_o}du\, e^{-u/{M'_M}^2}\left[\left(1-{m_c^4\over u^2}\right)
\left(1-{8u\over
m_c^2}+{u^2\over m_c^4}\right)-12\ln\left({m_c^2\over u}\right)\right]\right.
\nonumber \\*[7.2pt]
&+&\left.{\langle\overline{q}q\rangle^2\over6}e^{-m_c^2/{M'_M}^2}\right\} \; ,
\label{rslc}
\eeqa
\beq
|\lambda_N|^2=e^{M_N^2/M_M^2}\left({M_M^6\over32\pi^4}E_2+{2\over3}\langle
\overline{q}q\rangle^2\right)\; ,
\label{rsn}
\eeq
where $E_2=1-e^{-s_0/M_M^2}(1+s_0/M_M^2+s_0^2/(2M_M^4))$ accounts for the 
continuum contribution. For consistency we have also
neglected the contribution of the gluon condensate in the mass sum 
rules, since it is of little influence. 

In Eqs.~(\ref{rslc}) and (\ref{rsn})
$M_M^2$ and ${M'_M}^2$ represent the Borel masses in the two-point function
of the nucleon and $\Lambda_c$ respectively. Comparing Eqs.~(\ref{sr}),
(\ref{rslc}) and (\ref{rsn}) we can see that the exponentials multiplying
Eq.~(\ref{sr}) disappear if we choose
\beq
2M_M^2=M^2 \;\;\;\;\;\;\mbox{and}\;\;\;\;\;\;\;2{M'_M}^2=\mli\;.
\label{bo23}
\eeq
Indeed, this way of relating the Borel parameters in the two- and three-point
functions is a crucial ingredient for the incorporation of heavy quark
symmetries, and leads to a considerable reduction of the sensitivity to input
parameters, such as continuum thresholds $s_0$ and $u_0$, and to radiative
corrections \cite{ra,bbg}.

The parameter values used in all calculations are  $m_c=1.5\,
\GeV$, $m_D=1.87\,\GeV$, $m_{D^*}=2.01\,\GeV$,$M_N=938\,\MeV$, 
$M_{\Lambda_c}=2.285 \,\GeV$, $f_D=170\,\MeV$ $f_{D^*}=240\,\MeV$ 
\cite{bel},
$\langle\overline{q}q\rangle\,=\,-(0.23)^3\,\GeV^3$.
We parametrize the continuum thresholds as
\beq
s_0=(M_N+\Delta_s)^2\;,\label{s0}
\eeq
and 
\beq
u_0=(M_{\Lambda_c}+\Delta_u)^2\;.\label{u0}
\eeq
The values of $u_0$ and $s_0$ are extracted from the two-point
function sum rules for $M_{\Lambda_c}$ and $M_N$ in Eqs.~(\ref{rslc}) and 
(\ref{rsn}) and the respective sum rules in the {\bf{1}} structure given in 
refs.~\cite{ioffe,dosch,rry}. We found a good stability for $M_{\Lambda_c}$ 
and $M_N$,  being able to reproduce the experimental values for the masses
in the Borel mass $M_M^2\sim 1\GeV^2$ and ${M'_M}^2\sim 6\GeV^2$, 
with $\Delta_s=0.7\GeV$ and $\Delta_u=0.6\GeV$, which are the values that 
we are going to use in the calculations.

To allow for different values of $ M^2$ and $\mli$ we take them proportional
to  the respective baryon masses. In this way we 
study the sum rule as a function of $M^2$ at a fixed ratio 
\beq
{M^2\over\mli}={M_N^2\over M_{\Lambda_c}^2}\;.
\label{bo}
\eeq

In Fig.~2 we show the behavior of the perturbative, quark condensate 
and four quark condensate contributions
to the form factor $g_{ND\Lambda_c}(Q^2)$ at $Q^2=0\,\GeV^2$ (since in this 
case the cut in the $t$ channel starts at $t\sim m_c^2$ and thus the Euclidian
region stretches up to that threshold) as a function
of the Borel mass $M^2$. We observe that the different contributions
add to give a very stable result as a function of the Borel mass. Since
the Borel masses in the two- and three-point functions are related by 
Eq.~(\ref{bo23}) and since $M_M^2\sim 1\GeV^2$, to study the $Q^2$ dependence
of the form factor we fix $M^2=2.5 \GeV^2$ where the perturbative contribution
is the dominant one. The behaviour of the curve for other $Q^2$ values
is similar, however, for $Q^2>1.5\GeV^2$ the perturbative contribution is
no longer the dominant one. Therefore, in Fig.~3 we show the $Q^2$ dependence
of the form factor in the region  $0\leq Q^2\leq1.5\GeV^2$ (dots) where we 
believe we can trust the QCDSR results. We see
that the form factor is practically constant in this region, showing a tiny
decrease. We can fit the QCDSR results with a monopole form, as can be seen 
by the solid line in Fig.~3, and we get $g_{ND\Lambda_c}(Q^2)=843/(102+Q^2)$
which corresponds to a cut-off of order of $10\GeV$, much bigger than the
values used in ref.~\cite{tho2}. If we vary the value of the Borel mass
used to extract the $Q^2$ dependence of the form factor we can even get
a result that shows a tiny increase in the considered region. Therefore, 
in view of the uncertainties
involved in the approach, we can say that the QCDSR give a constant value for
the form factor. Considering $20\%$ variation in the continuum thresholds
and $f_D$ varying in the interval $f_D=170\pm10\,\MeV$ \cite{bel} we get:
\beq
g_{ND\Lambda_c}(Q^2)=7.9\pm0.9\; ,
\eeq
in agreement with our previous estimate of the coupling constant \cite{nn}.

The same analysis can be done for $g_{ND^*\Lambda_c}(Q^2)$ and
in Fig.~4 we show the perturbative, quark condensate and four quark 
condensate contributions
to the form factor $g_{ND^*\Lambda_c}(Q^2)$ at $Q^2=0\,\GeV^2$ 
as a function of the Borel mass $M^2$. Since the OPE sides of both form 
factors differ only by a sign the aspect of Fig.~2 is very similar to Fig.~4.
In Fig.~5 we show the $Q^2$ dependence
of the form factor $g_{ND^*\Lambda_c}(Q^2)$, extracted at $M^2=2.5\GeV$, 
in the region  
$0\leq Q^2\leq1.5\GeV^2$ (dots). Again we observe
that the form factor is practically constant in this region, showing a tiny
decrease. Fitting the QCDSR results with a monopole form 
(solid line in Fig.~5) we get $g_{ND^*\Lambda_c}(Q^2)=-204/(26.2+Q^2)$
which corresponds to a cut-off of order of $5\GeV$, still much bigger than the
values used in ref.~\cite{tho2}. Despite the fact that for this particular 
choice of $M^2$, $s_0$ and $u_0$ we get a form factor with a smaller cut-off
than obtained for $g_{ND\Lambda_c}(Q^2)$, varying $M^2$ and the continuun
thresholds we still can get results that show a tiny increase in the 
considered region. Therefore, also in this case, due to the uncertainties
involved in the approach, we conclude that the QCDSR give a constant value for
the form factor. Considering $20\%$ variation in the continuum thresholds
and $f_{D^*}$ varying in the interval $f_{D^*}=240\pm20\,\MeV$ \cite{bel} 
we get:
\beq
g_{ND^*\Lambda_c}(Q^2)=-7.5\pm1.1\; ,
\eeq

It is important to mention that constant form factors in hadronic vertices
have already been found in the context of QCDSR. This is the case of the
$g_{DD^*\pi}$ form factor. This form factor has been studied in the
QCDSR approach for an off-shell $D^*$ \cite{bel,bbd,kho}, and for an off-shell
pion \cite{nos}. As a matter of fact, in refs.~\cite{bel,bbd,kho} the 
authors analyze the semileptonic $f_+(t)$ form factor defined by
\beq
\langle\pi(p_\pi)|\bar{u}\gamma_\mu c|D(p_D)\rangle=(p_\pi+p_D)_\mu f_+(t)
+(p_D-p_\pi)_\mu f_-(t)\;,
\label{f+}
\eeq
where $t=q^2=(p_D-p_\pi)^2$. Since the vector current 
$V_\mu=\bar{u}\gamma_\mu c$
has the same quantum numbers as the vector meson $D^*$, the same sum
rules studied in refs.~\cite{bel,bbd,kho} can be used to study the
hadronic form factor $g_{DD^*\pi}(t)$, for an off-shell $D^*$ meson. It is
only the phenomenological side of the sum rule that has to be modified to
allow for a coupling between the vector current and the vector meson:
 \beqa
\langle\pi(p_\pi)|V_\mu|D(p_D)\rangle&=&\langle0|V_\mu|\pi(-p_\pi)D(p_D)\rangle
\nonumber \\*[7.2pt]
&=&\langle0|V_\mu|D^*(q)\rangle{1\over t-m_{D^*}^2}\langle D^*(q)\pi(p_\pi)|
D(p_D)\rangle\;.
\label{rel}
\eeqa

The $g_{DD^*\pi}(t)$ form factor is defined by the strong amplitude
\beq
\langle D^*(q,\epsilon)\pi(p_\pi)|D(p_D)\rangle=g_{DD^*\pi}(p_D+p_\pi)^\alpha
\epsilon_\alpha\;.
\eeq
Therefore, using the the vacuum to vector meson transition amplitude
defined in terms of the vector meson decay constant $f_{D^*}$: 
\beq
\langle D^*(q,\epsilon)|V_\mu|0\rangle=m_{D^*}f_{D^*}\epsilon^*_\mu
\; ;
\label{fd*}
\eeq
we can rewrite Eq.~(\ref{rel}) as
\beq
\langle\pi(p_\pi)|V_\mu|D(p_D)\rangle=-{m_{D^*}f_{D^*}g_{DD^*\pi}(t)\over
t-m_{D^*}^2}\left((p_\pi+p_D)_\mu-{q_\mu(p_D^2-p_\pi^2)\over m_{D^*}^2}\right)
\;. \label{gf}
\eeq

Comparing Eqs.~(\ref{f+}) and (\ref{gf}) we imediately see that $g_{DD^*\pi}
(t)$ and $f_+(t)$ are related by

\beq
f_+(t)={m_{D^*}f_{D^*}g_{DD^*\pi}(t)\over m_{D^*}^2-t}\;.
\label{refi}
\eeq

The authors of refs.~\cite{bel,bbd} claim that the QCDSR results for $f_+(t)$
can be well fitted by a monopole form with a pole mass $m_{pol}=m_{D^*}$.
According with Eq.~(\ref{refi}) this implies a constant $g_{DD^*\pi}(t)$. More
recently the authors of \cite{kho}, tried to fit their QCDSR results with a 
double pole parametrization of the type
\beq
f_+(t)={f_+(0)\over (1-t/m_{D^*}^2)(1-\alpha t/m_{D^*}^2)}\;.
\eeq
They have obtained $\alpha=0.01{\begin{array}{l}
                             +0.11\\
                             -0.07
			    \end{array}}$ and, therefore, they have concluded
that $\alpha$ is consistent with zero, which means the complete dominance
of the vector meson pole to $f_+$ with a consequent prediction of a constant
hadronic form factor in the $DD^*\pi$ vertex.

It is interesting to notice that, due to the uncertainties in the value of
$\alpha$ given above, one can not even rule out a  $DD^*\pi$ form factor
that slightly growns in the Euclidian region, as mentioned above in the case
of $ND(D^*)\Lambda_c$.

A very different result for the form factor in the $DD^*\pi$ vertex was  
obtained in the case of an off-shell pion \cite{nos}. This form factor is
important in the processes described in ref.~\cite{ko2}, which could 
contribute to explain the enhacement in the production of dileptons 
of intermediate masses, also observed in the NA50 experiment.
In the case that the pion is off-shell, the form
factor shows a very pronounced $Q^2$ dependence, with $Q^2$ being the squared
momentum of the off-shell pion in the Euclidian region. This result might be
suggesting that the size of a hadronic vertex depends on which particle
is off-shell. If the off-shell particle is light, then the vertex is not point 
like. However, if the off-shell particle is heavy, then the vertex is point 
like with a consequent constant form factor. Our result for the  $ND(D^*)
\Lambda_c$ form factors corroborate this hypothesis.

In conclusion, in this work we have calculated the form factors
for the hadronic vertices $ND\Lambda_c$ and $ND^*\Lambda_c$ using
QCD sum rules. These form factors are important to evaluate the 
charmonium dissociation cross section by hadrons, in the framework of 
meson exchange models
based on effective lagrangians. In the construction of the sum rules we have
performed a double Borel transformation with respect to the nucleon and
$\Lambda_c$ momenta, and we have evaluated the form factors as a
function of the heavy meson momentum $Q^2$. We have studied the sum rules 
in the structures $\sigma^{\mu\nu}\gamma_5 p^\prime_\mu p_\nu$ for 
$g_{ND\Lambda_c}(Q^2)$ and
$\rlap{/}{p^\prime}\gamma_\mu\rlap{/}{p}$  for $g_{ND^*\Lambda_c}(Q^2)$.
In the studied $Q^2$ region,
our results are compatible with constant form factors in these vertices. 
Considering 20\% of variation in the continuum thresholds, and around 10\% of 
variation in the meson decay constants we got:
\beqa
g_{ND\Lambda_c}(Q^2)&=&7.9\pm0.9 \nonumber\\
g_{ND^*\Lambda_c}(Q^2)&=&-7.5\pm1.1\; .
\eeqa

It is important to stress that constant form factors where also found, in the
framework of QCDSR, in the vertex $D^*D\pi$ with an off shell $D^*$.
Off course for very large values of $Q^2$ the form factors should go to zero,
but, according to our results, this means that the values of the cut-offs
in these form factors should be much larger than $2 \GeV$, which is the
typical value used in refs.~\cite{ko,tho2,ha2}.

\underline{Acknowledgements}: This work has been supported by FAPESP
and CNPq.

\begin{figure} \label{fig1}
\begin{center}
\vskip -1cm
\epsfysize=7.0cm
\epsffile{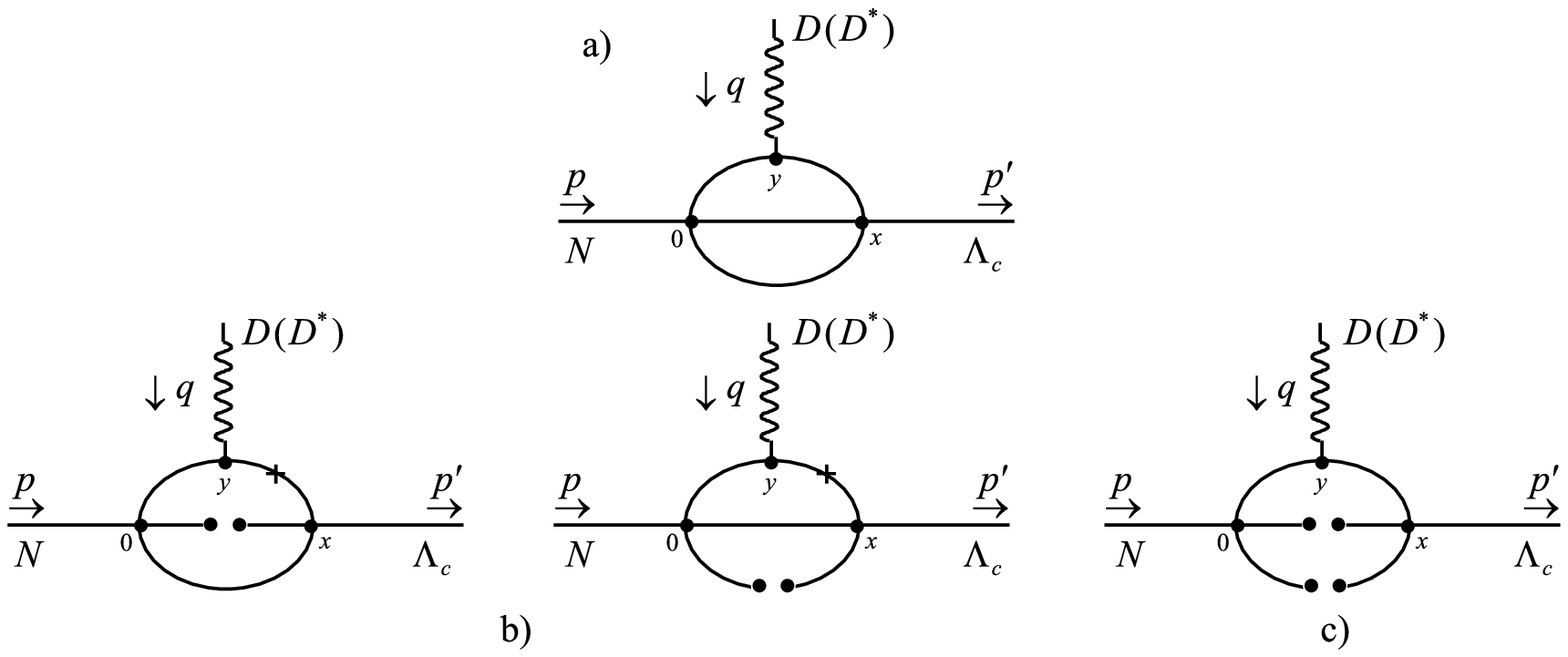}
\caption{Diagrams that contribute to the OPE side of $g_{ND(D^*)\Lambda_c}
(Q^2)$. The cross stands for a charm mass insertion.}
\end{center}
\end{figure}

\begin{figure} \label{fig2}
\leavevmode
\begin{center}
\vskip -1cm
\epsfysize=9.0cm
\epsffile{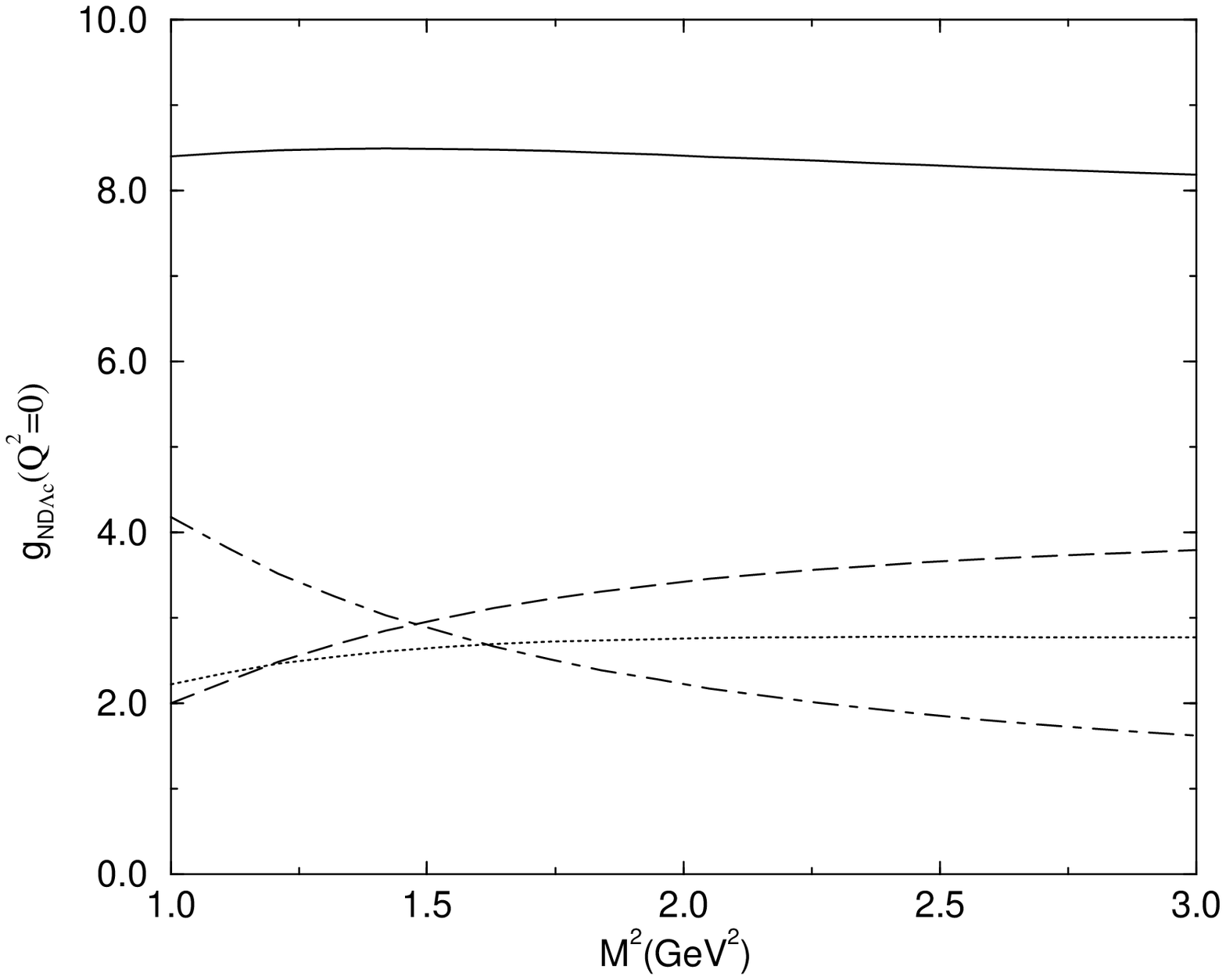}
\caption{Borel dependence of the perturbative (dashed line), quark condensate
(dotted line) and four quark condensate (dot-dashed line) contributions
to the $ND\Lambda_c$ form factor (solid line) at $Q^2=0$.}
\end{center}
\end{figure}

\begin{figure} \label{fig3}
\leavevmode
\begin{center}
\vskip -1cm
\epsfysize=9.0cm
\epsffile{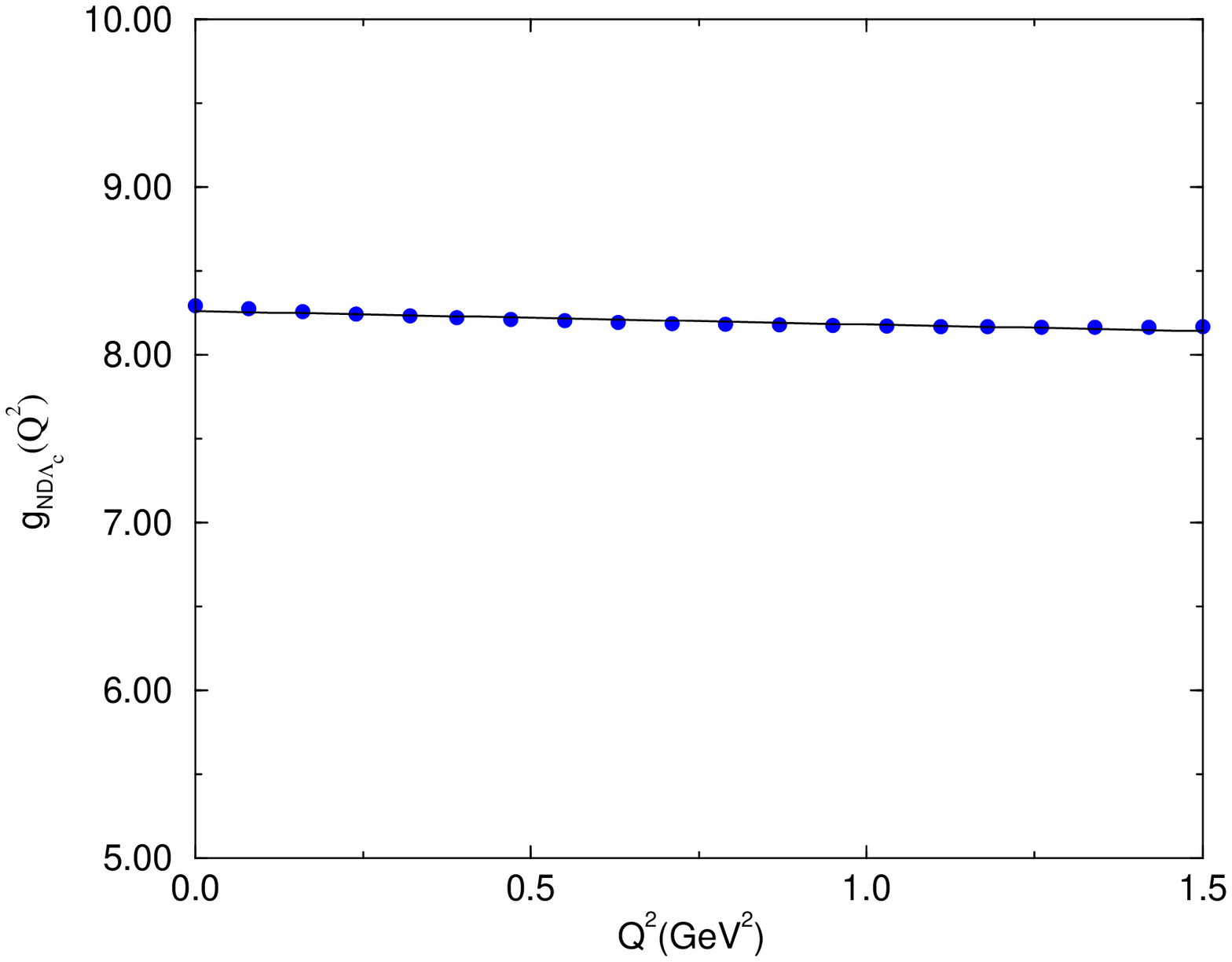}
\caption{Momentum dependence of the $ND\Lambda_c$ form factor 
(dots). The solid line give the parametrization of the QCDSR results with
a monopole form.}
\end{center}
\end{figure}

\begin{figure} \label{fig4}
\leavevmode
\begin{center}
\vskip -1cm
\epsfysize=9.0cm
\epsffile{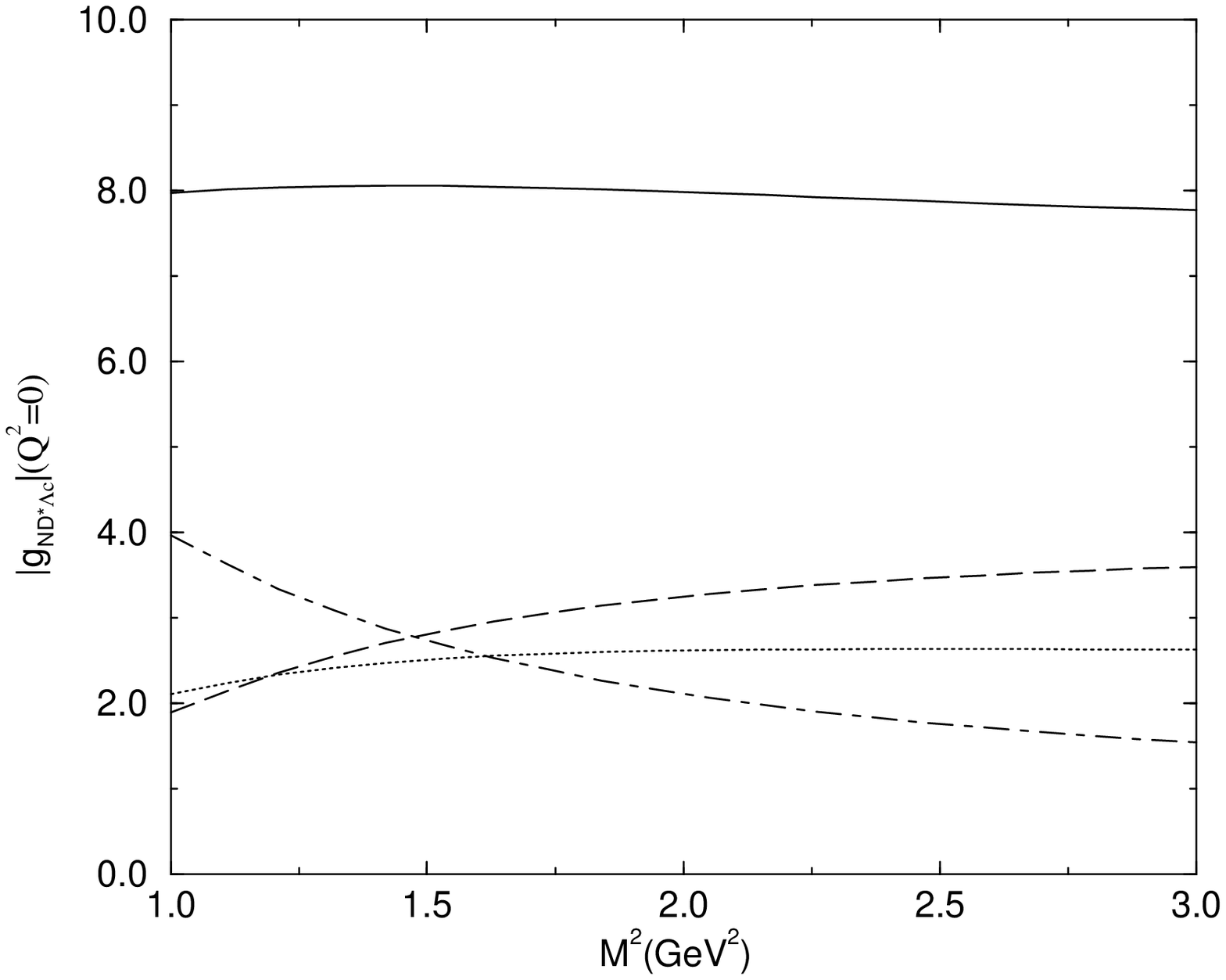}
\caption{Borel dependence of the perturbative (dashed line), quark condensate
(dotted line) and four quark condensate (dot-dashed line) contributions
to the $ND^*\Lambda_c$ form factor (solid line) at $Q^2=0$.}
\end{center}
\end{figure}

\begin{figure} \label{fig5}
\leavevmode
\begin{center}
\vskip -1cm
\epsfysize=9.0cm
\epsffile{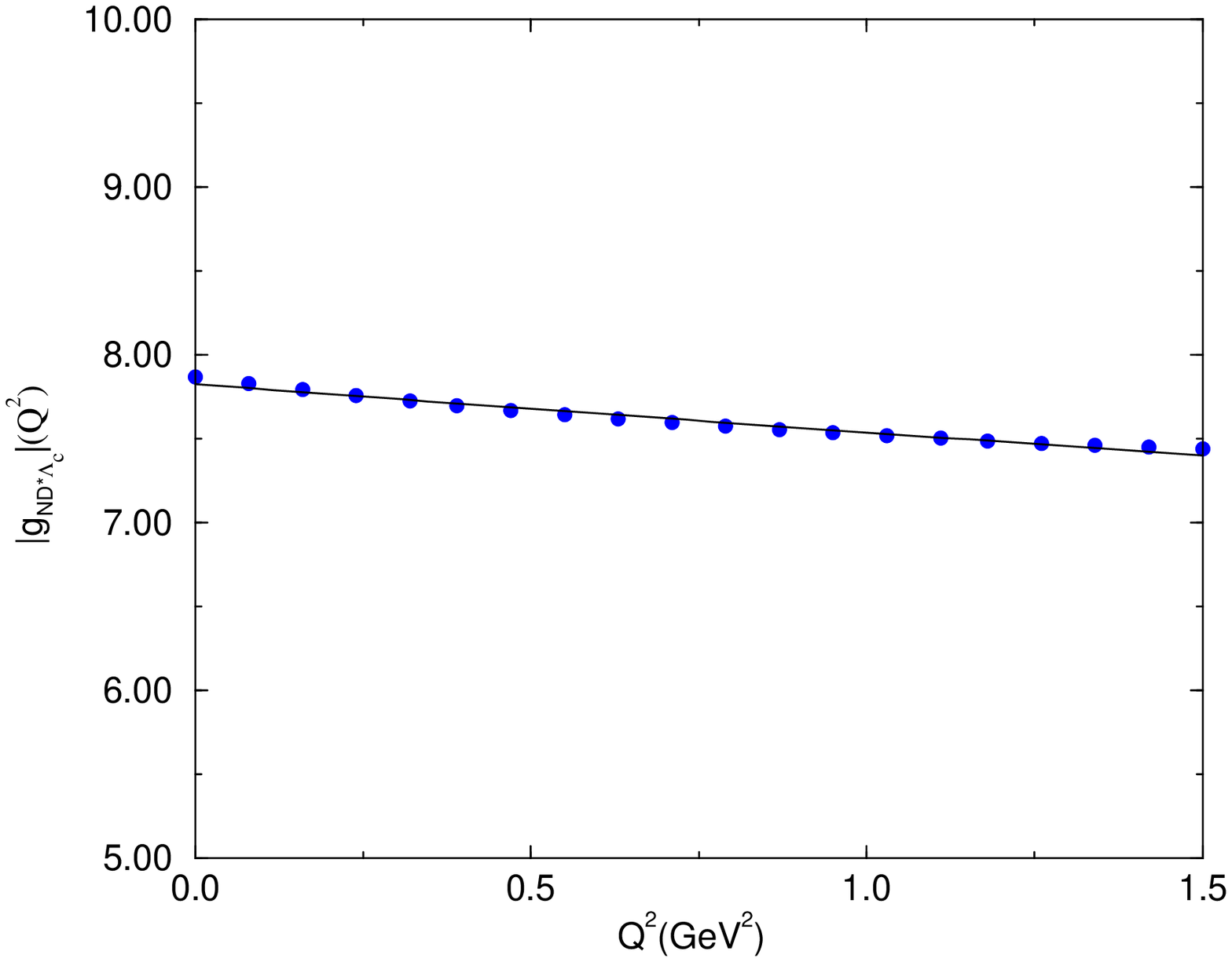}
\caption{Momentum dependence of the $ND^*\Lambda_c$ form factor 
(dots). The solid line give the parametrization of the QCDSR results with
a monopole form.}
\end{center}
\end{figure}


\vspace{0.5cm}


\begin{thebibliography}{99}



\bibitem{NA50} NA50 Collaboration (M.C. Abreu et al.), {\sl Phys. Lett.}
{\bf B477}, 28 (2000).

\bibitem{nu01} W. Cassing, E.L. Bratkovskaya and S. Juchem, nucl-th/0001024,
 {\sl Nucl. Phys.} {\bf A674}, 249 (2000).

\bibitem{he02} A. Capella, E.G. Ferreiro and A.B. Kaidalov, hep-ph/0002300,
{\sl Phys. Rev. Lett.} {\bf85}, 2080 (2000).

\bibitem{tho1} A. Sibirtsev, K. Tsushima, K. Saito and A.W. Thomas, 
nucl-th/9904015,  {\sl Phys. Lett.} {\bf B484}, 23 (2000).

\bibitem{mamu} S.C. Matinyan and B. M\"uller, nucl-th/9806027, 
{\sl Phys. Rev.} {\bf C58}, 2994 (1998). 

\bibitem{ha} K.L. Haglin, nucl-th/9907034, {\sl Phys. Rev.} {\bf C61}, 
031902 (2000). 

\bibitem{ko} Z. Lin and C.M. Ko, nucl-th/9912046, {\sl Phys. Rev.} {\bf C62}, 
034903 (2000). 

\bibitem{tho2} A. Sibirtsev, K. Tsushima and A.W. Thomas, nucl-th/0005041.

\bibitem{ha2} K.L. Haglin and C. Gale, nucl-th/0010017.

\bibitem{svz}  M.A. Shifman, A.I. Vainshtein and V.I. Zakharov, {\sl Nucl. 
              Phys.}  {\bf B120}, 316 (1977). 


\bibitem{nn}   F.S. Navarra and M. Nielsen, 
               {\sl Phys. Lett.} {\bf B443}, 285 (1998).

\bibitem{ioffe} B.L. Ioffe, {\sl Nucl. Phys.} {\bf B188}, 317 (1981).

\bibitem{dosch} E. Bagan, M. Chabab, H.G. Dosch and S. Narison,
                {\sl Phys. Lett.} {\bf B278}, 367 (1992);
		{\bf B287}, 176 (1992); {\bf B301}, 
                243 (1993).


\bibitem{ra} R.S. Marques de Carvalho et al., {\sl Phys. Rev.} {\bf D60}, 
034009 (1999).


\bibitem{klo} H. Kim, S.H. Lee and M. Oka, {\sl Phys. Lett.} {\bf B453}, 199 
(1999); {\sl Phys. Rev.} {\bf D60}, 034007 (1999);  H. Kim, T. Doi, M. Oka
and S.H. Lee, nucl-th/0002011.

\bibitem{bnn} M.E. Bracco, F.S. Navarra and M. Nielsen, 
               {\sl Phys. Lett.} {\bf B454}, 346 (1999).

\bibitem{io2}  B.L. Ioffe and A.V. Smilga, {\sl Nucl. Phys.} {\bf B216} 373
(1983); {\sl Phys. Lett.} {\bf B114}, 353 (1982).

\bibitem{cut} R.E. Cutkosky, {\sl J. Math Phys.} {\bf1}, 429 (1960).

\bibitem{rry}  L.J. Reinders, H. Rubinstein and S. Yazaki, {\sl Phys.
               Rep.} {\bf 127}, 1 (1985). 

\bibitem{bbg} E. Bagan, P. Ball and P. Gosdzinsky, {\sl Phys. Lett.} 
{\bf B301}, 249 (1993). 

\bibitem{bel} V.M. Belyaev, V.M. Braun, A. Khodjamirian  and R. R\"uckl,
 {\sl Phys. Rev.} {\bf D51}, 6177 (1995). 

\bibitem{bbd} P. Ball, V.M. Braun and H.G. Dosch, {\sl Phys. Lett.} {\bf B273},
316 (1991).

\bibitem{kho} A. Khodjamirian et al., hep-ph/0001297, to appear in 
{\sl Phys. Rev. D}.

\bibitem{nos} F.S. Navarra, M. Nielsen, M.E. Bracco, M. Chiapparini and
C.L. Schat, hep-ph/0005026, {\sl Phys. Lett.}  {\bf B489},  319  (2000). 

\bibitem{ko2} Z. Lin, C.M. Ko and B. Zhang, {\sl Phys. Rev.} {\bf C61}, 
024904 (2000). 



\end{thebibliography}
\end{document}